\documentclass[12pt]{article}
\usepackage{graphicx}
\usepackage{psfrag}
\usepackage{epstopdf}
\usepackage{latexsym}

\begin{document}

\begin{titlepage}
\null\vspace{-62pt}

\pagestyle{empty}
\begin{center}

\vspace{1.0truein} {\Large\bf Neutrino oscillations in gravitational and cosmological backgrounds}

\vspace{1in}
{\large George Koutsoumbas and Dimitrios Metaxas*} \\
\vskip .4in
{\it Department of Physics,\\
National Technical University of Athens,\\
Zografou Campus, GR 15773 Athens, Greece\\
kutsubas@central.ntua.gr, metaxas@central.ntua.gr*}\\

\vspace{.5in}
\centerline{\bf Abstract}

\baselineskip 18pt
\end{center}
We use the eikonal approximation in order to calculate the additional phase shift between two neutrino mass eigenstates during their propagation in a background of gravitational wave or scalar perturbations in the flat and the FRW spacetime metric. We comment on the dependence of the results on the characteristics of the perturbations, give some order-of-magnitude estimates, and find that, although small, the resulting phase difference persists for large redshifts, up to the validity of our approximations.

\end{titlepage}
\newpage
\pagestyle{plain}
\setcounter{page}{1}
\newpage

\section{Introduction}

The phenomenon of flavor oscillations because of the difference between mass and flavor neutrino eigenstates is well-known and described by quantum field theory and quantum mechanics in terms of the phase difference of the respective wave functions during their propagation in flat spacetime \cite{neutrinos1}.  Similar calculations have been done in curved spacetime backgrounds \cite{neutrinos2} and are of great theoretical interest, although the gravitational effects are generally too small to be observed experimentally.

Because of the various experimental developments in neutrino physics, and related astrophysical and cosmological observations \cite{neutrinosexp}, as well as the possible relevance of similar calculations in other early universe processes \cite{lepto}, we consider here the oscillations between two neutrino flavors in backgrounds involving scalar and tensor perturbations for the flat and the FRW metric.

We start by describing in Sec.~2 the eikonal (WKB) approximation used, first in flat spacetime and then its generalization in curved spacetime, in order to set up our formalism 
and make contact with previous works. As a first application, we consider a gravitational wave perturbation in a flat background and give some order-of-magnitude estimates
for the phase difference, finding an interesting dependence on the frequency of the perturbation and the energy of the respective particles (a similar problem has recently been considered in \cite{maxim} but with a different geometry involved).

 In Sec.~3 we consider scalar and tensor perturbations in Friedmann-Robertson-Walker (FRW) spacetime, and apply our previous formalism in order to calculate the resulting
phase difference and its dependence on the parameters of the problem. We find that, although for a pure FRW spacetime the phase difference tends to a constant value for large 
redshifts \cite{visi}, the additional phase difference due to the perturbations persists and provides, therefore, an interesting result regarding bursts from the early universe.

In Sec.~4 we conclude with some comments, relations with other similar works, and possible directions for future work.

\section{Eikonal approximation and a gravitational wave background}

In flat spacetime, with metric $(+---)$, the eikonal (WKB) approximation to the scalar, Klein-Gordon equation,
\begin{equation}
(\partial^2 + m^2) {\sf \Phi }=0,
\label{scalar1}
\end{equation}
is obtained by writing the scalar field, with mass $m$, as ${\sf\Phi} ={\sf\Phi_0} \, e^{iS}$, in terms of a slowly varying amplitude, ${\sf\Phi_0}$, and a phase, $S$. After expanding and considering the various resulting terms one gets the eikonal equation,
\begin{equation}
\partial_{\mu} S \,\partial^{\mu} S = m^2,
\label{scalar2}
\end{equation}
if the approximations
$\partial^2 S << (\partial S)^2$ and $\partial_\mu {\sf\Phi_0} << {\sf\Phi_0} \, \partial_\mu S$ are satisfied.

A similar expansion can be made for a fermion field, which also satisfies the Klein-Gordon equation, since $(\gamma^\mu \partial_\mu)^2 = \partial^2$ for the Dirac matrices,
$\gamma^\mu$.

For a freely moving particle, either a scalar or a fermion, one gets,
\begin{equation}
S=-E t + k_1 x_1 + k_2 x_2 + k_3 x_3,
\end{equation}
with the respective energy and momenta satisfying $E^2 - {\vec{k}}^2 = m^2$.

As mentioned in the Introduction, we will consider two species of neutrinos,  propagating along the $x_3$-axis, and write
\begin{equation}
S=- E t + K x_3,\,\,\,{\rm with}\,\,\, K=(E^2 - m^2)^{1/2} \approx E -\frac{m^2}{2 E} -\frac{m^4}{8 E^3},
\label{expansion1}
\end{equation}
for almost massless, highly relativistic particles. Then, the standard calculation \cite{neutrinos1} gives the result for the oscillation probability
between, for example, the electron and muon neutrinos, $\nu_e$ and $\nu_\mu$, with a mixing angle, $\theta$,
emitted at $(t_1, x_{31})$ and detected at $(t_2, x_{32})$, as
\begin{equation}
P(\nu_e \rightarrow \nu_\mu) = \sin^2(2\theta) \, \sin^2 \left(\frac{\delta \phi_{12}}{2}\right),
\label{oscex}
\end{equation}
with
\begin{equation}
\delta \phi_{12} = - \delta S = \frac{\delta m^2}{2 E} L + \frac{\delta m^4}{8 E^3} L,
\label{formula1}
\end{equation}
where $L=x_{32}-x_{31} \approx T=t_2 -t_1$, and $\delta m^2 = m_1^2 -m_2^2,\, \delta m^4 = m_1^4 -m_2^4$ for the different mass eigenstates.
The oscillation probability, (\ref{oscex}), as well as the survival probability, $P(\nu_e \rightarrow \nu_e) =1-P(\nu_e \rightarrow \nu_\mu) $,
can be measured experimentally, and the physical  parameters $\delta m^2$ and $\theta$ (or its generalization from the PMNS matrix for three generations)
can be determined. The phase difference, $\delta \phi$ (or $\delta S$), although not directly observable, is an important factor, that enters in 
(\ref{oscex}) (and its generalization), for the consistency of the calculation.

In curved spacetime, with metric tensor $g_{\mu\nu}$, $g=\det{g_{\mu\nu}}$, Riemann tensor $R^\alpha_{\beta \gamma \delta} = \partial_\gamma \Gamma^\alpha_{\beta \delta} - \cdots$,
and covariant derivative, $\nabla_\mu$, 
because of the Lichnerowicz formula \cite{spin},
 $(\gamma^\mu \nabla_\mu)^2 = \Box  -R/4$, where $R$ is the Ricci scalar and
 $\Box = g^{\mu\nu} \nabla_\mu \nabla_\nu$, one gets from  the Dirac equation for a fermion field ${\sf\Psi}$, the corresponding generalised equation
\begin{equation}
\left(\Box + m^2 - \frac{R}{4}\right){\sf \Psi }=0,
\end{equation}
and the eikonal approximation 
for an approximately constant spinor, with only a phase factor with spacetime dependence,
${\sf\Psi} = {\sf\Psi_0} \, e^{iS}$, then gives
\begin{equation}
g^{\mu\nu} \partial_\mu S \, \partial_\nu S = m^2 -\frac{R}{4}
\label{fermion1}
\end{equation}
provided that the conditions $g^{\mu\nu} \partial_\mu\partial_\nu S = \partial^2 S << g^{\mu\nu} \partial_\mu S \partial_\nu S = (\partial S)^2$
and $\partial_\mu (\sqrt{-g} g^{\mu\nu}) << \sqrt{-g} g^{\mu \nu} \partial_\mu S$ are satisfied. The presence of the curvature term in (\ref{fermion1}) is essentially due to the Lichnerowicz 
formula, and derived here in the eikonal approximation for an approximately constant spinor, with only a varying phase factor; it should be noted, however, that some works in \cite{neutrinos2} consider a covariantly constant spinor and then separate the exponential phase factor in a physical phase and a phase associated with the connection terms, thereby eliminating the curvature term in (\ref{fermion1}). The difference in the oscillation probabilities between the two approaches comes from the second order terms (of order $\delta m^4$) in (\ref{formula1}), and will be subleading in the cases and redhifts we consider here, as we discuss in the next Section. We keep it in our formulas, however, since in problems with stronger gravity contributions or higher redshifts this term may become important (interestingly, a similar curvature term arises for the effective Hamiltonian in the path integral quantization of a general dynamical system in curved spacetime \cite{dewitt}).

The curved spacetime result for the phase difference has been studied in various works and for various gravitational backgrounds \cite{neutrinos2, maxim, visi, pop, maxim2}.
The result for the gravitational contribution to the phase difference is generally too small to be detected, since, however, there are several searches involving 
various orders of magnitude of the related parameters, and because of the theoretical importance of these and similar considerations, 
it is useful to examine the problem in various physically significant situations.
The contribution of the curvature term in  (\ref{fermion1}) is generally also too small to be significant for current measuments, even compared to other contributions
related to the cosmological expansion, since, however, we are also interested in possible extensions of the formalism in other primordial problems, we will keep it in the
formulas derived in the next Section.  As mentioned before, it is possible that a more complete treatment of the fermion propagation in these backgrounds would be desirable, 
as well as a search for some exactly solvable cases, in order to determine the contributions of strong curvature terms.

Here, as a first application of the formalism developed, we will start with the consideration of the background of a plane gravitational wave in 
flat spacetime (a similar problem, but with a different geometry, has also been studied in \cite{maxim}).
The gravitational wave is assumed to have a plane geometry,
 also propagating along the $x_3$-axis, with
\begin{equation}
ds^2 = g_{\mu\nu} dx^\mu dx^\nu =
dt^2 -  g_{ij}dx^i dx^j,
\end{equation}
where
\begin{equation}
g_{ij} = \delta_{ij} - h_{ij} =
\left(
\matrix{
   1+h_+ & h_\times & 0   \cr
    h_\times & 1 - h_+ &  0   \cr
   0 & 0 & 1 }.
\right)
\label{gwm}
\end{equation}

We will consider the polarization $h_\times =0, h_+ = c_+ \, e^{i\omega (x_3-t)}$,  and a similar treatment for the other polarization is obvious.
In this background we have $R=0$ at this order and (\ref{fermion1}) becomes
\begin{equation}
\left( \frac{\partial S}{\partial t} \right)^2 - (1-h_+) \left( \frac{\partial S}{\partial x} \right)^2 -
(1+h_+) \left( \frac{\partial S}{\partial y} \right)^2 - \left( \frac{\partial S}{\partial z} \right)^2 = m^2
\end{equation}
The fermions will also be assumed to propagate in the $x_3$-direction, but with a possible small transverse momentum component,
and we write
\begin{equation}
S=-E t + k_1 x_1 + k_2 x_2 + K  x_3 + \alpha_+ e^{i\omega (x_3-t)},
\end{equation}
in order to solve for $\alpha_+$, and get
\begin{equation}
2 \alpha_+ i \omega (E-K)=c_+ (k_2^2 -k_1^2).
\label{pop1}
\end{equation}
Writing 
\begin{equation}
k_\perp^2 = k_1^2 + k_2^2 << K^2 , \,\, \, k_+^2 = k_1^2 - k_2^2, \,\,\,
K \approx E - \frac{k_\perp^2 + m^2}{ 2 E},
\label{pop2}
\end{equation}
we find
\begin{equation}
\alpha_+ = \frac{i c_+}{\omega} \frac{k_+^2}{k_\perp^2 + m^2} E
\label{formula2}
\end{equation}
for the additional contribution to the phase of the particles propagating in the gravitational wave background.

In order to get an order of magnitude estimate for the effect, we consider two terrestrial detectors, situated at a transverse distance $X$ apart (in the $x_1$-direction),
and a neutrino point source at a distance $L$ (in the $x_3$-direction).
The first detector, assumed exactly at the $x_3$-direction, observes particles without any transverse momentum component, and the oscillation phase for these is given
by the original formula (\ref{formula1}). The particles arriving at the second detector have an additional phase difference, given by their transverse motion in the $x_1$-direction
and the contribution of the gravitational wave background.
For $k_2=0,  \, k_1\approx E \frac{X}{L}, \,  k_1^2<<m^2 $, and $k_1  X$ much smaller than the other phase factors involved,
 this extra contribution to the oscillation phase is
\begin{equation}
 \delta \tilde{\phi}_{12} \approx \frac{k_1^2 \, \delta m^2}{m^4}\frac{E}{\omega} A_{gw}
\approx \frac{E^3}{\omega} \frac{X^2}{L^2} \frac{\delta m^2}{m^4} A_{gw},
\label{rev1}
\end{equation}
where $A_{gw}$ is the magnitude of the gravitational wave perturbation.

This functional dependence of the phase difference in (\ref{rev1}) should be compared with the usual result for flat spacetime (\ref{formula1}), as well as the result for 
the gravitational correction to the phase difference for neutrinos emitted in radial distance $r_1$ and detected at $r_2$,  in the Schwarzchild background of a mass $M$, $\delta\phi_{12SC}=\frac{\delta m^4}{8 E^3} 2GM \log\frac{r_1}{r_2}$, where $G$ is Newton's constant \cite{neutrinos2}. Since (\ref{rev1}) increases with the particle energy, one gets different results for the 
respective oscillation lengths and a possibly larger contribution, depending on the order of magnitude estimates.

It should also be mentioned that our result gives this enhancement only for neutrinos moving at approximately the same direction as the gravitational wave.
This may seem as a special case, but since various experimental searches are investigating multi-messenger, high energy astrophysical signals \cite{mm}, it concerns a limit that is quite relevant  phenomenologically. In the case of neutrinos moving opposite to the wave, with a factor $E+K$ instead of $E-K$  in (\ref{pop1}), one obviously does not get a similar kind of enhancement for the phase difference. The eikonal equation in the gravitational wave background was previously solved in \cite{pop} and considered in \cite{maxim2} but for different physical problems and geometries (for example, in \cite{maxim2} an average of the signal over all directions of a stochastic gravitational wave background was performed, thereby not considering the enhancement shown here).

Using some rough values for the relevant parameters, $X\sim 10^3 km, \, L\sim 10^8 pc, \, \omega\sim 100Hz, \, A_{gw}\sim10^{-21} $,
and taking as parameters the masses and energies of the observed neutrinos,
$\delta m^2 \sim m^2 \sim 10^{-w} eV, E\sim 10^n eV$, we can write the order of magnitude of the additional phase difference in (\ref{rev1}) as
$\delta \tilde{\phi}_{12} \sim 10^{3 n + w - 47}$. 
A more concrete estimate, using $\delta m^2 \sim 10^{-3}-10^{-4}  eV, m^2\sim 1 eV$ \cite{exp1},
gives, for the same astrophysical source, a phase difference of the order of unity for neutrino energies at the range of $ E \sim 10^5 TeV$.
It is interesting that
such energies are, in principle, accessible by the
modern neutrino telescopes like the IceCube and ANTARES collaborations, which can detect ultra-high energy
neutrinos with energies up to $ EeV = 10^6 TeV$
\cite{eev}.

It should noted, however, that, although the phase difference for the neutrino paths may become of order unity and therefore, in principle, observable with these values,
it is highly unlikely that both events of a gravitational wave burst and neutrino emission would occur at the same time interval, since they are both transient with a small 
time window of a few milliseconds
for the usual astrophysical processes.
If the neutrino beam and the gravitational wave come from strictly the same source then they both travel in the same direction and the phase difference is zero.
The presence of two distinct sources at comparable astrophysical coordinates, operating  in the same time window, is, of course, highly improbable. It is interesting, however,
that the order of magnitude of a purely gravitational effect is, in principle, within the present experimental limits. It is also possible that our results can be used
in conjunction with other geometries, as in \cite{maxim2}, in order to further explore similar effects.

Because of the large astrophysical distances involved, one should also consider the problem of the eventual decoherence of the neutrino beam \cite{decoherence}.
We find that the bound
\begin{equation}
L\stackrel{<}{\sim} L_{\rm coh}  \sim \frac{E^2}{\sigma_p \, \delta m^2}
\label{decoherence}
\end{equation}
is satisfied for the above indicative values of the relevant parameters for $\sigma_p \stackrel{<}{\sim} 10^5 eV$.
($L_{\rm coh}$ is the coherence length and $\sigma_p$ is the momentum dispersion of a neutrino beam with energy $E$ \cite{decoherence}).

The above are, of course, some initial crude estimates;
 we considered a gravitational wave  and neutrinos originating from two distinct but overlapping astrophysical sources,
and it is meant only as an indication of the order of
magnitude of the relevant gravitational contribution (which, interestingly, is
proportional to $E/\omega$ instead of inversely proportional to $E$ as in (\ref{formula1}) and other cases).

\section{Tensor and scalar perturbations in FRW spacetime}

Since various high-energy astrophysical phenomena are associated with higher redshift values, it is of interest to extend the previous results to an expanding, FRW background, with metric
\begin{equation}
ds^2 = dt^2 -a^2(t)\,d\vec{x}^2=a^2(\eta)(d\eta^2 - d\vec{x}^2),
\end{equation}
where we consider the flat case, and frequently use the conformal time, $\eta$, instead of the physical time, $t$,
since $\eta$ is directly related to the comoving distance for highly relativistic 
particles moving along null rays.

Primes and dots will denote derivatives with respect to $\eta$ and $t$ respectively, and the Hubble expansion rate,
$H=\dot{a}/a =a'/a^2$, satisfies
\begin{equation}
H^2 = H_0^2 \left( \frac{\Omega_r}{a^4} +\frac{\Omega_m}{a^3} +\Omega_\Lambda \right),
\label{cosm1}
\end{equation}
with $H_0$ its present value and $\Omega_r +\Omega_m + \Omega_\Lambda =1$.
Here, $\Omega_r$, $\Omega_m$ and $\Omega_\Lambda$ are the ratios of the radiation, matter and vacuum energy densities to the 
critical density, $3 H_0^2/8\pi G$.
The value of the Ricci scalar is given by
$R=-\frac{6 a''}{a^2} = -3 H_0^2 \left(\frac{\Omega_m}{a^3} +4 \Omega_\Lambda \right)$,
and we will also use as a parameter the redshift, $z$, with
$1+z = \frac{1}{a}$ and $d\eta=-\frac{dz}{H(z)}$.

We consider highly relativistic neutrinos emitted at $\eta_i$ with redshift $z_i$,
and received at the present time, $\eta_0$, with $a(\eta_0)=1$ and redshift $z_0=0$,
having traveled the comoving distance
$\eta_0-\eta_i = \int_0^{z_i}\frac{dz}{H(z)} = \Delta\eta = \Delta x_3 = L$.

The eikonal equation, (\ref{fermion1}),  after writing
\begin{equation}
S=-S_0(\eta) + K x_3,
\end{equation}
becomes
\begin{equation}
\left( \frac{\partial S_0}{\partial \eta} \right)^2 = K^2 + a^2(\eta)\left( m^2 - \frac{R}{4}\right)
\end{equation}
and gives
\begin{equation}
S_0(\eta_0) =\int_{\eta_i}^{\eta_0} \left[ K^2 + a^2 (\eta) (m^2 +3\, \Omega_\Lambda H_0^2)
+\frac{3}{4} \frac{H_0^2 \,\Omega_m}{a(\eta)}\right]^{1/2}  d \eta
\end{equation}
or, in terms of the redshift,
\begin{equation}
S_0(z_i) = \int_0^{z_i}  \frac{dz}{H(z)} f(z)
\label{zeq}
\end{equation}
where
\begin{equation}
f(z, m^2)=\left[ K^2 + (m^2 +3 \,\Omega_\Lambda H_0^2)(1+z)^{-2}
+\frac{3}{4} H_0^2\,\Omega_m(1+z) \right]^{1/2}.
\label{zeq1}
\end{equation}

The particle energy in the eikonal approximation is given by
\begin{equation}
E=\frac{\partial S}{\partial t} =\frac{1}{a}\frac{\partial S_0}{\partial\eta_0}
\end{equation}
or
\begin{equation}
 E^2 =\frac{K^2}{a^2} +(m^2+3 \Omega_\Lambda H_0^2)+\frac{3}{4}\frac{H_0^2 \Omega_m}{a^3},
\end{equation}
and the energy of the fermions observed at present time, with $a=1$, is given by
\begin{equation}
E_0^2 = K^2 +\tilde{m}^2
\end{equation}
with 
$\tilde{m}^2 = m^2 + c_0 H_0^2$, where
$c_0 =3\Omega_\Lambda +\frac{3}{4}\Omega_m$, and accordingly we get
$\delta \tilde{m}^2=\delta m^2$, $\delta\tilde{m}^4 = \delta m^4 + 2 c_0 H_0^2 \,\delta m^2$,
and expansions similar with (\ref{expansion1}).

For particle energies and momenta much higher than the other scales involved,
specifically for $K^2>> m^2, \, H_0^2 \,(1+z)$, 
and for very small redshifts, such that $z\approx 0, \, a\approx 1$,
$E_0=\partial S_0/\partial\eta$,
we can write $S_0 \approx \frac{\partial S_0}{\partial \eta} \Delta\eta$,
and use the previous approximations, 
$K\approx E_0 -\frac{\tilde{m}^2}{2 E_0} -\frac{\tilde{m}^4}{8 E_0^3}$,
in order
 to get the phase difference between two fermion species as
\begin{equation}
\delta\phi_{12}= L\left( \frac{\delta m^2}{2 E_0} + \frac{\delta m^4}{8 E_0^3}
+\frac{c_0 H_0^2\, \delta m^2}{4 E_0^3} \right),
\label{cf1}
\end{equation}
which agrees with the flat spacetime result (\ref{expansion1}) in the case of zero expansion rate. The last term, which is a subleading correction for the relevant values of $H_0^2$ and $\delta m^2$, is the contribution of the curvature term in
(\ref{fermion1}), that results from the fermion propagation in curved spacetime.

The phase difference can also be calculated from the above expressions for arbitrary redshift \cite{visi}.
After expanding
\begin{equation}
\delta\phi_{12}= \frac{\partial S_0}{\partial m^2}\delta m^2 + \frac{1}{2}\frac{\partial^2 S_0}{\partial (m^2)^2}\delta m^4+\cdots,
\label{zeqq}
\end{equation}
and keeping the previous approximation for  $K^2>> m^2, \, H_0^2 \,(1+z)$, but for finite redshifts, $z$,
we get 
\begin{equation}
\delta\phi_{12}=\frac{\delta m^2}{2 E_0} F_1(z_i)+\frac{\delta m^4}{8 E_0^3} F_2(z_i)+\cdots,
\label{zfig}
\end{equation}
with the functions $F_1, F_2, ...$, calculated from (\ref{zeq}), (\ref{cosm1}), (\ref{zeqq}),
\begin{equation}
F_1(z)=\frac{1}{H_0}\int_0^z \frac{dz}{(1+z)^2 g(z)}\,\,,\,\,\,
F_2(z)=\frac{1}{H_0}\int_0^z \frac{dz}{(1+z)^4 g(z)}\,\,,
\label{visi2}
\end{equation}
with
$g(z)=\left[\Omega_\Lambda +\Omega_m (1+z)^3 \right]^{1/2}$,
and we plot the first term, $F_1(z)$,  in the first graph of Fig.~1.
We see that
for small values of the redshift, $z\approx 0$, $F_1(z)$ is equal to the physical distance $L\approx z/H_0$, and the 
result for the phase difference agrees with the flat spacetime expression (\ref{expansion1}).
For larger redshifts it tends to a constant value, and we have a similar behavior for $F_2$,
so neutrino oscillations do not differentiate between different sources of large $z$,
this conclusion will be modified, however, when we consider metric perturbations.

Now we can examine perturbations in the FRW metric, and we start with the tensor perturbations, with metric
\begin{equation}
ds^2 = a^2(\eta) \left[ d\eta^2 -(\delta_{ij} - h_{ij}) dx^i dx^j \right],
\end{equation}
where $h_{ij}$ has a similar form as (\ref{gwm}).
Then, for the  $+$ polarisation,
\begin{equation}
h_+ =\frac{c_+}{a} e^{i\omega (x_3 -\eta)}
\end{equation}
(the scale dependence corresponds to perturbations with $\omega^2 >> a''/a$)
we write
\begin{equation}
S=-S_0(\eta) + k_1 x_1 + k_2 x_2 + K x_3 + \tilde{\alpha}_+ e^{i \omega (x_3 -\eta)},
\end{equation}
where $S_0(\eta)$ is given by the previous result for the FRW metric without the perturbation
(we use the same notations for the $k_i$ as in the flat spacetime case)
and apply
(\ref{fermion1}) 
to get
\begin{equation}
2 \tilde{\alpha}_+ \left( \frac{\partial S_0}{\partial \eta} - K \right) = \frac{i c_+}{a\omega} k_+^2.
\end{equation}
For the phase of the neutrinos observed at the present time, with $a=1$ and $K>>m^2$,
we have $E_0=\partial S_0/\partial\eta$ and
after expanding,
$\frac{\partial S_0}{\partial \eta} - K \approx \frac{\tilde{m}^2 + k_\perp^2}{2 E_0}$, we get
\begin{equation}
\tilde{\alpha}_+ \approx \frac{i c_+ k_+^2 E_0}{\omega \, (\tilde{m}^2+k_\perp^2)},
\label{formula3}
\end{equation}
a result  similar to (\ref{formula2}), that reduces to it in the limit of zero expansion.
As far as the various orders of magnitude involved are concerned, the same comments apply here
as in the flat case, and similar conclusions are reached when one considers the phase difference
of neutrinos propagating in a cosmological gravitational wave background.
We note, however, that the amplitude of the oscillating phase, $\tilde{\alpha}_+$, does not vary with 
the scale factor, so the result for the additional phase difference does not tend to a constant for large $z$,
like the first terms decribed before.

We can also consider scalar perturbations, with metric
\begin{equation}
ds^2 = a^2 (\eta) \left[ (1 + 2 \Phi) d \eta^2 -(1-2 \Phi) d \vec{x}^2 \right].
\end{equation}
A more general analysis can be easily made for perturbations that are both space and time dependent,
since, however, we are interested here in late-time applications, we consider time independent
perturbations with $\Phi= c_s \,e^{i \vec{k} \cdot \vec{x}}$, and, after writing
\begin{equation}
S= -S_0 (\eta) + K x_3 + \alpha_s \, e^{i k x_3},
\end{equation}
we use (\ref{fermion1}) to get
\begin{equation}
\alpha_s = \frac{i c_s}{k}\,\frac{\left(\frac{\partial S_0}{\partial\eta}\right)^2 + K^2}{K}.
\label{ss1}
\end{equation}
For observations at present time with $a=1$, we have $E_0 =\partial S_0/\partial\eta$, and 
after expanding, as usual,
$K\approx E_0 -\frac{\tilde{m}^2}{2 E_0} -\frac{\tilde{m}^4}{8 E_0^3}$,  the second fraction in 
(\ref{ss1})
becomes $2 E_0 +\frac{\tilde{m}^4}{4 E_0^3}$, so we get for the phase difference between the two neutrino species
an additional factor, given by
\begin{equation}
\delta\phi_s = \frac{c_s}{i k}\frac{\delta \tilde{m}^4}{4 E_0^3} \, e^{i k x_3}.
\label{ss2}
\end{equation}
The result is of higher order than the other contributions, in terms of $\delta m^2$ and $E_0$, it is, however, inversely
proportional to the wavenumber of the perturbation and increasing for large wavelengths. We also note that, like the previous
result for the tensor perturbations, it does not become constant for larger redshifts as opposed to the case in \cite{visi, silk}.

In the second graph of Fig.~1, we show this result added to the term for the phase difference calculated from (\ref{zeq})
that is proportional to $\delta m^4/8 E_0^3$,
that is, $F_2(z)$ of (\ref{zfig}).
In the graphs in Fig.~1 the vertical axis is in units of $H_0^{-1}$
and the horizontal axis is the redshift, $z$.
We used $\Omega_m=0.3$ and $\Omega_\Lambda=0.7$ for simplicity, and in the second graph we 
used the result from (\ref{ss2}) that corresponds to a wavelength $10^{-3} H_0^{-1}$,
with an amplitude of the scalar perturbation, $c_s$ equal to $10^{-3}$ (we used this value in the Figure instead of a realistic order of magnitude of $10^{-5}-10^{-6}$   in order
to demonstrate the final effect). Since the effect of the curvature term here is also subdominant, we have not included it (that is, we take
essentially $\delta \tilde{m}^4 \approx \delta m^4$) we have kept this distinction in the formulas, however, since similar calculations may be relevant to
considerations of physical problems in primordial situations.

We also note that, if one considers, for example, the black hole metric as a perturbation to the flat spacetime (for appropriately large distances)
one gets from (\ref{ss2}) (in the limit of zero expansion) the expressions derived before in the literature \cite{neutrinos2}, so our formulas are consistent with other existing results.

As  far as the tensor perturbations in an FRWl background are concerned, the same order of magnitude estimates are involved as in the previous Section,
and we see a substantial enhancement in the phase difference for neutrinos travelling in approximately the same direction as the gravitational wave. 
Again, although these estimates are interestng, it is highly unlikely that two such separate sources of neutrino and gravitational wave bursts would happen to 
occur in the same small time window.
For the propagation in a scalar perturbation background the result is of the same order of magnitude as previously derived expressions in gravitational backgrounds
such as the Schwarchild metric and therefore too small to be of phenomenological relevance.
It is also interesting, however, that for both scalar and tensor perturbations the phase difference persists even for large redshifts.

\section{Comments}

In this work we derived expressions and estimates for the phase difference associated with fermion (neutrino) oscillations
in gravitational and cosmological backgrounds, and showed that, although subleading,
the corrections may be interesting depending on the characteristics of the scalar and tensor perturbations involved,
and also that they persist for larger values of the redshifts.

Our results are consistent with previously derived expressions and provide an additional formalism that may be applied
to other problems of interest in cosmology and neutrino physics. We also gave some order-of-magnitude estimates
for the corrections, in order to examine the possibility of their significance in other cosmological and astrophysical problems.

In future works, it would be interesting to consider the gravitational effects on more than two fermion generations, as well as
similar physical problems in primordial situations and effects of stronger curvature, modified gravities, and other high-energy 
astrophysical processes \cite{mod}.

\begin{figure}
\centering
\includegraphics[width=90mm]{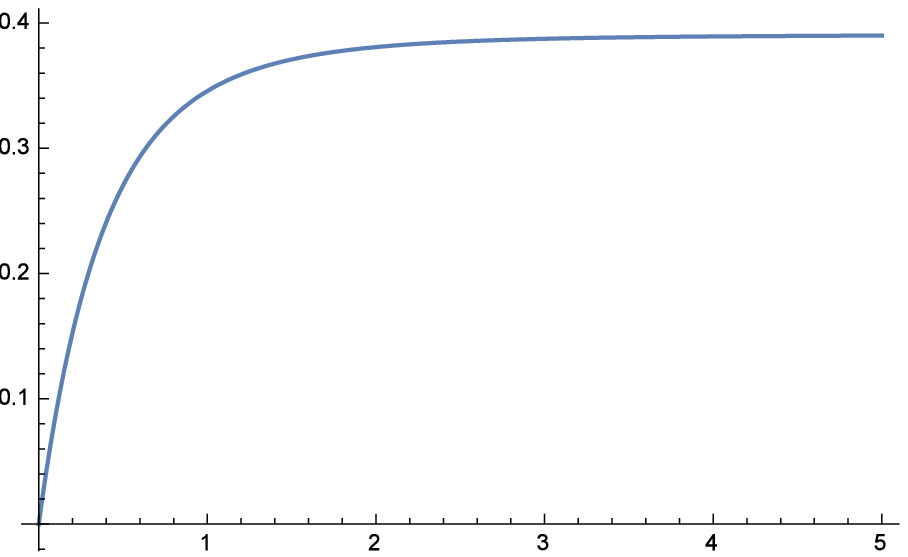}
\includegraphics[width=90mm]{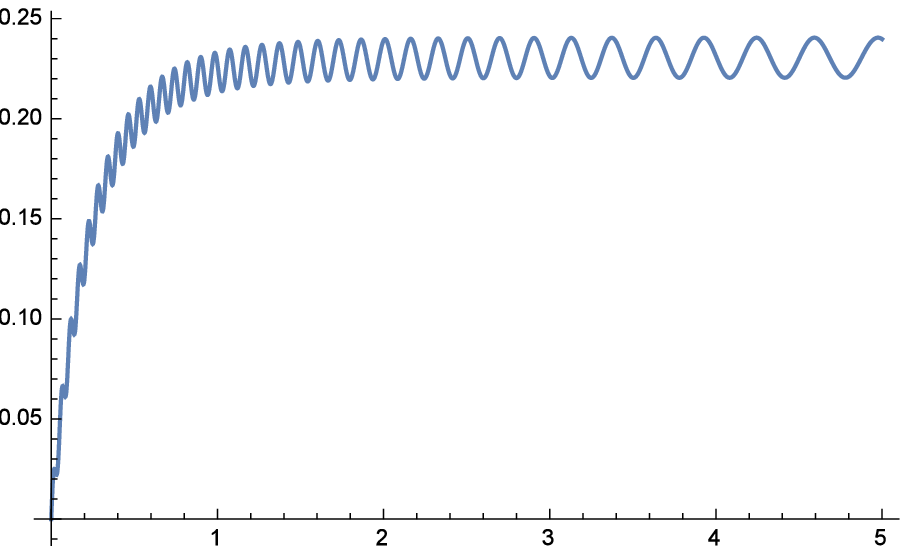}
\caption{ We show the results for the phase difference in FRW spacetime with scalar perturbations as functions of the redshift, $z$.
The first graph is $F_1(z)$ of (\ref{zfig}), that is the result for the phase difference that is proportional to $\delta m^2/2 E_0$ 
(which is not affected by the perturbations).
The second graph shows the result for the phase difference
that is proportional to $\delta m^4 /8 E_0^3$, that is, $F_2(z)$ of (\ref{zfig}), with the contribution from (\ref{ss2}) added.
The vertical axis is in units of $H_0^{-1}$ for both graphs.  Other parameters involved are described in the text following (\ref{ss2}).
We note, for example, that we used the value of $10^{-3}$ for the scalar perturbations, instead of a more realistic value of $10^{-6}$, in order to
demonstrate the effect in the Figure.
 }
\end{figure}

\end{document}